\documentclass{aa}
\usepackage{epsfig}
\newcommand{\Ha}{H$\alpha$}

\newcommand{\Hb}{H$\beta$}
\newcommand{\Hbm}{\rm H\beta}
\newcommand{\oi}{{\rm \mbox{[O\,\sc i]}}}
\newcommand{\oii}{{\rm \mbox{[O\,\sc ii]}}}
\newcommand{\oiii}{{\rm \mbox{[O\,\sc iii]}}}
\newcommand{\nii}{{\rm \mbox{[N\,\sc ii]}}}
\newcommand{\neiii}{{\rm \mbox{[Ne\,\sc iii]}}}
\newcommand{\sii}{{\rm \mbox{[S\,\sc ii]}}}
\newcommand{\niioiii}{{\rm \mbox{[N\,\sc ii]}}/{{\rm \mbox{[O\,\sc iii]}}}}
\newcommand{\ohb}{{\rm \mbox{[O\,\sc iii]/{H$\beta$}}}}
\newcommand{\oha}{{\rm \mbox{[O\,\sc i]/{H$\alpha$}}}}
\newcommand{\sha}{{\rm \mbox{[S\,\sc ii]/{H$\alpha$}}}}
\newcommand{\hahb}{H$\alpha$/H$\beta$}
\newcommand{\nha}{{\rm \mbox{[N\,\sc ii]/H$\alpha$}}}

\newcommand{\hii}{{H\,\sc ii}}
\newcommand{\etal}{{\rm et al.}\ }
\begin{document}
\thesaurus{6(11.09.4;11.19.2;;11.19.3; 11.01.1;11.11.1;09:08.1)}
\title{Star formation in bar environments\thanks{Observations reported
    in this paper were obtained at the Multiple Mirror Telescope
    Observatory, a facility operated jointly by the University of
    Arizona and the Smithsonian Institution}}
\subtitle{II. Physical properties, age and abundances of H\,{\sc ii} regions}
\author{P. Martin \inst1 \and D. Friedli \inst2} 
\offprints{P. Martin} 
\institute{
Canada-France-Hawaii Telescope, PO Box 1597, Kamuela, HI 96743, USA. 
E-mail: martin@cfht.hawaii.edu 
\and
Geneva Observatory, CH-1290 Sauverny, Switzerland. 
E-mail: Daniel.Friedli@obs.unige.ch}
\date{Received: March 15, 1999; accepted: April 8, 1999} 
\maketitle
\markboth{P. Martin \& D. Friedli: Star formation in bar environments. II}{}

%-----------------------------
\begin{abstract}
  The nebular properties (electronic density, extinction, age, O/H
  abundances) of \hii\ regions found along the bars of the sample of
  barred spiral galaxies studied by Martin \& Friedli (1997) are
  examined.  From line ratio diagnostic diagrams, it is showed that
  regions located along the major axis of the bars have a normal
  photoionization spectrum, that is, line ratios reproductible from
  nebular conditions and ionizing star radiation field normally
  encountered in extragalactic \hii\ regions. There is an indication,
  however, that their degree of ionization might be somewhat
  different. Another ionization mechanism (high-velocity shocks or
  hard UV radiation) is clearly present for regions found nearby the
  centers of the galaxies. The electronic density of the regions along
  the bars is very close to that of disc regions ($\langle$
  $N_e$$\rangle$ $\sim$ 80 cm$^{-3}$).  On average, bar and disc
  regions have a similar visual extinction ($A_V \sim 1$\,mag) with
  exceptions for some regions located near the bar dust lanes of the
  earlier types of galaxies in our sample.  Although the average \Ha\ 
  equivalent width of bar \hii\ regions ($\sim$250\,{\AA}) is half
  that of disc regions, this disparity could be due to uncertainties
  in the galactic continuum and does not translate into a significant
  age difference. The oxygen abundance distribution was also
  investigated in the bar of these galaxies.  The O/H scatter was
  found to be very small ($<$0.1\,dex) indicating that mixing of the
  chemical composition by gas flows is very efficient in a barred
  structure.

\end{abstract}
%________________________________________________________________

%----------------------------------------------------------------------------
%
%   INTRODUCTION
%
\section{Introduction}
%-----------------------------------------------------------
Galactic bars are the sites of highly diversified star formation
activities.  Phillips (1993, 1996), Garc\'{\i}a-Barreto \etal (1996)
and Martin \& Friedli (1997, hereafter MF97) have all showed that {\it
  along} certain bars, mostly found in late-type spirals, star
formation (SF) can be quite intense.  For instance, the bar of the
SBcd galaxy NGC\,4731 has a total star formation rate (SFR) of about
1.5 M$_\odot$ yr$^{-1}$ (MF97).  However, in other cases, SF in the
bar can be very weak or completely absent. This is the case for most
of the bars in early-type barred spirals (e.g. NGC\,1300, NGC\,1512,
NGC\,3351).  The origin of these differences is not yet completely
understood.  Numerical simulations (Friedli \& Benz 1995, MF97,
Martinet \& Friedli 1997) and observations (Martin \& Roy 1995)
suggest that the existence of massive star formation in certain bars
is a relatively brief event ($\sim$0.5--1.0\,Gyr) in the evolution of
a barred system and that it mostly takes place during the formation of
the bar itself.  The amplitude and duration of the event is, however,
controlled by a complicated combination of different physical
processes.  For example, the time-dependent bar evolution, initial gas
content, amplitude of gas flows, and mechanical energy injected in the
interstellar medium (ISM) by supernovae ejecta are all factors that
can influence the level of SF activity in bars (MF97).  Hence, it is
essential to acquire more data on the properties of the \hii\ regions
formed in such environments to better constrain the relative
importance of these factors, and consequently significantly improve SF
recipes used in numerical simulations.
  
The morphology, SFRs and other properties of SF along the bars of a
sample of eleven spiral galaxies were studied in the first paper in
this series (MF97).  Including the large diversity in SF activity,
MF97 found that the distribution of \hii\ regions can be highly
asymmetrical in the bar. For some SBc spirals, large regions can be
present outside the bar major axis.  This morphology suggests that the
SF process along the bar of late-type spirals is a chaotic process,
not strictly confined along the major stellar/gas orbits defined by
the barred potential.  In such a case, one could expect similar
properties for these \hii\ regions when compared to the disc star
forming regions.  The situation could be different for bars in earlier
types of galaxies for which the star forming regions are generally
located next to dust lanes (MF97). For one galaxy of their sample
(NGC\,7479), MF97 also estimated the amount of gas flowing in the bar
and falling into the galaxy center.  They found that possibly as much
as 75\% of the gas in the bar is not transformed into stars.
Numerical simulations suggest that this number is very dependent on
the presence of bar-induced shocks in the star forming ISM.  It is
then critical to determine whether there is some signature of these
shocks in the \hii\ regions along the bars.

In this paper, spectrophotometric data are used to derive the physical
properties of a sub-sample of \hii\ regions found along the bars of
the galaxies studied by MF97.  As a comparison sample, a few regions
located in the discs of these galaxies have also been analysed.  Using
different diagnostic line ratio diagrams as defined by Baldwin \etal
(1981) and Veilleux \& Osterbrock (1987), the excitation of regions
located in bar environments is compared to that of ``normal'' disc
regions (Sect.~3.1).  A similar analysis is performed for the
electronic density using the appropriate sulfur line ratio
(Sect.~3.2).  The distributions of the visual extinction for both
populations are also studied in Sect.~3.3. Using the \Ha\ equivalent
width indicator (e.g. Leitherer \etal 1999), we also infer the
approximate age of \hii\ regions (Sect.~3.4).  Since large-scale
mixing of the chemical composition by bar-driven gas radial flows
occurs in galaxy discs (Martin \& Roy 1994; Friedli \etal 1994), the
O/H distribution in bar environments is also investigated (Sect.~3.5).

%---------------------------------------------------------------------
%
%  DATABASE 
%
\section{Database}
%-----------------------------------------------------------
\subsection{Observations}
The long slit observations were conducted during two runs in April and
October 1994 at the equivalent 4.5 meter Multi-Mirror Telescope on Mt
Hopkins, Arizona.  The ``Blue Channel'' spectrograph was used with a
Loral 3k$\times$1k CCD. A grating with 500 grooves/mm and a blaze at
5410\,{\AA} was employed; the spectral range covered in the reduced
spectra is about 3500\,{\AA} ($\sim$3500 $\rightarrow$ $\sim$7000)
with a dispersion of 1.17\,{\AA}/pixel. An order sorting filter was
used (UV-36) and all the data were obtained with the CCD binned by a
factor of two in the spatial direction (0.6\arcsec/pixel).  All these
observations were carried out with an unvignetted
2\arcsec$\times$150\arcsec\ slit positioned close to the parallactic
angle to avoid any light loss. Three exposures of 15 or 20 minutes
were obtained for each slit position in the bar.  Between exposures,
the alignment of the MMT six mirrors was verified and corrected, if
necessary.  Numerous standard stars were observed with a larger slit
(5\arcsec) during the night for the flux calibration procedure.

Table~1 presents the journal of observations for the \hii\ regions of
ten of the eleven objects studied by MF97 (NGC\,5068 was not
observed).  For some objects, a few slit positions were required to
optimize the number of \hii\ regions observed in the bar and the disc.
Average seeing during these observations was about 1.2\arcsec\ and
conditions were photometric.  Note that the only galaxy in the sample
considered as an AGN (LINER) is NGC\,7479.

%-----------------------------------------------------------
\begin{table*}[t]
\caption{Journal of observations.}
\begin{tabular}{llccccccc} 
\hline\hline
Galaxy & Epoch & \# Slits & Slit PA [$^\circ$] & Exposure Times [s] 
& $<$Airmass$>$ & $N_{\rm bar}$$^a$ & $N_{\rm disc}$$^a$ & L$_{bar}$$^b$ [kpc]
\smallskip
\\ \hline
NGC\,1073 & 1994 Oct 9  & 1 &  68 & 3 $\times$ 1200 & 1.2 & 2 & 2 & 5.0\\
NGC\,1087 & 1994 Oct 9  & 1 & 139 & 3 $\times$ 1200 & 1.4 & 2 & 2 & 1.6\\
NGC\,3319 & 1994 Apr 17 & 1 &  40 & 1 $\times$ 1200 & 1.0 & 2 & 3 & 4.0\\ 
NGC\,3359 & 1994 Apr 16 & 1 &  26 & 3 $\times$ 1200 & 1.2 & 8 & 5 & 5.8\\
NGC\,3504 & 1994 Apr 16 & 2 & 140; 175 & 2 $\times$ 3 $\times$ 900 & 1.1 & 6 & 7 & 6.8\\
NGC\,4731 & 1994 Apr 16 & 1 & 126 & 3 $\times$  900 & 1.3 & 6 & 4 & 15.2\\
NGC\,4900 & 1994 Apr 16 & 1 & 145 & 3 $\times$  900 & 1.3 & 4 & 6 & 3.4 \\
NGC\,5921 & 1994 Apr 16 & 1 & 159 & 3 $\times$ 1200 & 1.2 & 5 & 4 & 8.1\\
NGC\,7479 & 1994 Oct 4  & 2 &   2; 11 & 2 $\times$ 3 $\times$ 1200 & 1.3 & 7 & 9 & 16.6\\
NGC\,7741 & 1994 Oct 4  & 2 &  92; 98 & 2 $\times$ 3 $\times$ 1200 & 1.1 & 7 & 1 & 3.5\\ \hline
\end{tabular}
\begin{minipage}{160mm}
\smallskip
$^a$ Number of bar and disc \hii\ regions with detected \Hb.\\
$^b$ Total length of the bar based on distance and measurements given in MF97.
\end{minipage}
\end{table*}

%-----------------------------------------------------------
\subsection{Data reduction and analysis}
The long slit spectra were reduced following standard procedures
available in the {\sc longslit} package in {\sc iraf} {\footnote{IRAF
    is distributed by the National Optical Astronomy Observatories,
    which are operated by the Association of Universities for Research
    in Astronomy, Inc., under cooperative agreement with the National
    Science Foundation}. First, a bias and flat-field correction was applied. 
The illumination pattern along the slit was corrected using a set of sky
flat-fields taken in the same optical configuration.  Wavelength
calibration was done using a Helium--Neon--Argon exposure taken
immediately after the science observation.  Geometric distortion and
alignment of the spectra were corrected by 2D-mapping of the spectral
lines from the calibration sources using the {\sc fitcoords} and {\sc
  transform} algorithms.
  
The next step, sky subtraction, was done by extracting a sky
background using sections along the long slit outside the galaxy.  To
make sure that the signal from the galaxy was minimized, the average
sky backgrounds were compared between the different slit positions
since most of the slits used for the disc \hii\ regions were less
contaminated by the disc emission.  The spectra were then flux
calibrated using a sensitivity function obtained from a set of
standard stars observed through the nights.  Extraction of the
spectrum for each \hii\ region was done using the spatial profile at
H$\alpha$ seen along slit.  In general, no continuum trace was found
for these \hii\ regions.  The extraction trace was forced across the
spectral range using the positions of the main nebular lines along the
spectral domain.  Finally, the individual spectra were combined.

The integrated fluxes of the main nebular lines \Hb,
\oii\,$\lambda$3727, \oiii\,$\lambda\lambda$4959,\,5007,
\oi\,$\lambda$6300, \nii\,$\lambda$6584, \Ha, and
\sii\,$\lambda\lambda$6717,\,6731 were measured using a Gaussian
fitting algorithm available with {\sc splot} in {\sc iraf}.  A
background continuum estimated from each side of the lines was
automatically subtracted.

Many \hii\ regions in bars show a strong underlying Balmer absorption
at \Hb.  McCall \etal (1985) have shown that adding about 2\,{\AA} of
equivalent width for normal disc regions constitutes an appropriate
correction.  This correction was applied for all the \Hb\ fluxes for
the \hii\ regions in our control sample and the bar regions with
shallow underlying absorption.  However, for about 20\% of the bar
sample, this correction was not sufficient. For these regions, mostly
located in the bars with a strong continuum (e.g. NGC 3504, NGC 5921,
NGC 7479), the amplitude of the underlying absorption was estimated
from a Gaussian fit.  In general, we found that about 5\,{\AA} of
equivalent width were necessary to assure a good correction.  This
value was applied for all bar regions with absorption higher than
normal.

All of the line fluxes were corrected for interstellar reddening by
comparing the \hahb\ ratio to the theoretical Balmer decrement (2.86)
for Case B recombination (that is, for nebulae with large optical
depths for the H\,{\sc i} resonance lines) at $10^4$\,K. In reality,
the temperature of bar \hii\ regions is probably around 7000--8000\,K
due to their high O/H abundances (see below) so that our extinction
based on the Balmer decrement might be overestimated by about
0.1\,mag.  Since the Balmer ratio represents only an approximation of
the real extinction, we did not take this difference into account.
The reddening law formulated by Savage \& Mathis (1979) was assumed
for reddening correction.

It is generally difficult to evaluate the accuracy of the absolute
spectrophotometric fluxes for individual objects.  The main
uncertainties include the contamination by the bright galactic
continuum, the underlying absorption at \Hb, the accuracy of the
photometric calibration, the spectrum extraction in a crowded field of
\hii\ regions, the positioning of the slit, the correction for the
interstellar extinction and, of course, the intensity of the lines.
We evaluate the accuracy of the fluxes to be at about 20--30\%. Since
we will only compare the global behavior of \hii\ regions, these high
uncertainties should not influence our conclusions.

%-----------------------------------------------------------
%
% SPECTROPHOTOMETRIC SECTION
%  
\section{Nebular properties}
%
%-----------------------------------------------------------
\subsection{Excitation}
The basic source of ionization in an \hii\ region is the UV radiation
field from young massive stars.  The emission line spectra resulting
from a pure photoionization field in a Str\"omgren sphere can be
studied using diagnostic diagrams (Baldwin \etal 1981; Evans \& Dopita
1985; Veilleux \& Osterbrock 1987; Osterbrock 1989; Dopita \&
Sutherland 1995; Rola et al. 1997).  These standard diagrams are based
on nebular line ratios like \ohb, \oha, \nha, and \sha.  For high
redshift galaxies, the red part of the spectrum is shifted to the
near-infrared. If optical studies are needed, other diagnostic lines
like \oii\,$\lambda$3727 or \neiii\,$\lambda$3869 can also be relied
upon (Rola \etal 1997).  These diagnostic diagrams are extremely
useful in distinguishing between normal photoionized regions and
regions with another ionization mechanism (e.g. high-velocity shocks,
hard UV fields).  The line ratios above are also independent of the
reddening correction and depend only on the accuracy achieved for the
spectrophotometry.  However, despite the usefulness of these
diagnostic diagrams for determining whether another ionization
mechanism is present or not, it is very difficult to identify this
mechanism.  For this, sophisticated nebular models must be used (e.g.
Stasi\'nska 1990; Dopita \& Sutherland 1995).

%-----------------------------------------------------------
\begin{figure}[t]
\vskip -0.3truecm
\centerline{
\psfig{figure=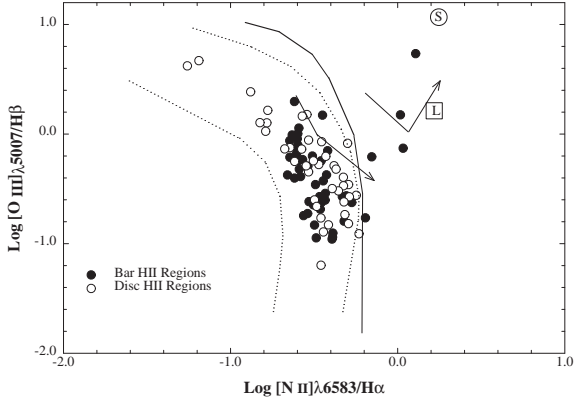,width=8.8cm,clip=}
}
\vskip -0.5truecm
\caption[]{Diagnostic diagram of our sample of bar and disc \hii\
  regions based on Osterbrock (1989).  The sequence of normal regions
  is represented by the dotted lines.  The full line indicates the
  separation between normal regions and regions with another
  ionization mechanism.  The squared and circular symbols correspond
  to the average location of Seyfert~2 and LINER galaxies from Rola
  \etal (1997).  The arrow lines indicate the effect of high-velocity
  shocks and magnetic fields on the line ratios from models by Dopita
  \& Sutherland (1995).  From the beginning to the end of the arrow,
  the velocity for the shock varies from 150 to 500\,km\,s$^{-1}$. The
  magnetic field parameter is zero for the bottom arrow and
  4\,$\mu$G\,cm$^{3/2}$ for the top arrow}
\end{figure}    

%-----------------------------------------------------------
Figure~1 illustrates the first diagnostic diagram with a correlation
between \ohb\ and \nha\ for the bar regions (dark symbols) and the
disc regions of our control sample (open symbols).  The dotted lines
define the limits of the sequence of normal \hii\ regions that can be
found in Osterbrock (1989) and Kennicutt \etal (1989).  The full line
is the separation between normal photoionized regions and regions with
another ionization mechanism.  We have also indicated the effect of
high-velocity shocks and magnetic fields on these line ratios from
models by Dopita \& Sutherland (1995). The average location of Seyfert
2 and LINER galaxies in this diagram are also displayed. At this time,
it is still not entirely clear that the spectral characteristics
(mainly the high \nha\ ratio) of these last objects are due to high UV
radiation or high-velocity shocks (see discussion by Dopita \&
Sutherland 1995). However, Fig. 1 shows that excepting for four
regions, {\it all the \hii\ regions in our sample are located inside
  the sequence of normal \hii\ regions}.  These four regions are
located close to the nucleus in NGC\,3504 (a starburst galaxy) and
NGC\,7479 (LINER).  Thus, from this diagram alone, bar \hii\ regions
do not exhibit any sign of high-velocity shocks or hard-UV radiation.
Apart from these central peculiarities, there is no dependence on
radius.

Although \nha\ is a good diagnostic ratio for high-velocity shocks or
very hard ultraviolet radiation (Veilleux \& Osterbrock 1987; Dopita
\& Sutherland 1995), other ratios like \sha\ or \oha\ are more
sensitive to these ionization mechanisms.  Figure~2 illustrates
another diagnostic diagram: the correlation between \ohb\ and \sha.
The dashed and full lines are as in Fig.~1.  Arrows also show the
effect of the high-velocity shocks and magnetic fields.  Although most
of the \hii\ regions fall within the normal region sequence, there is
a small number of bar regions outside the sequence.  Figure~3 shows
the sequence between \ohb\ and \oha. The latter ratio was detected in
about 67\% of the sample of bar regions but only in about 40\% of the
disc region sample.

From these diagrams, it is clear that {\it most bar \hii\ regions do
  not exhibit any systematic evidence of high-velocity shocks
  ($>$150\,km\,s$^{-1}$) or very hard UV radiation}. Only
circumnuclear regions exhibit obvious signs of another ionization
mechanism (fast shocks and/or hard-UV radiation), as expected from the
work of Kennicutt et al. (1989). Nevertheless, if these conditions do
not appear to be present in bar regions, Figs. 2 and 3 suggest that
shocks with lower velocity or an abnormal UV photoionization field
cannot be excluded.  This possibility is discussed in Sect.~4.

%-----------------------------------------------------------
\begin{figure}[t]
\vskip -0.3truecm
\centerline{
\psfig{figure=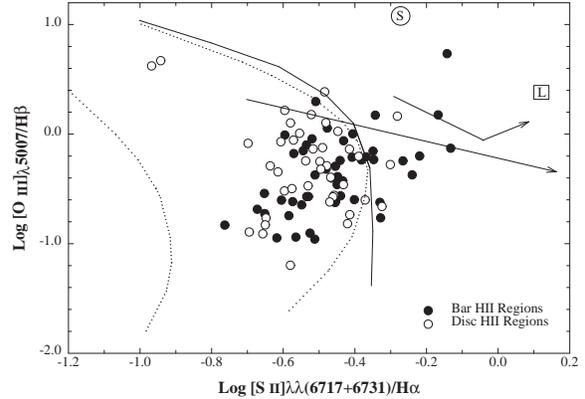,width=8.8cm,clip=}
}
\vskip -0.5truecm
\caption[]{Diagnostic diagram comparing bar and disc \hii\ regions.
  The dotted lines are the sequence of normal regions defined by
  Osterbrock (1989) and the full line separates the normal regions
  from regions which are not exclusively photoionized.  The square and
  circular symbols correspond to the average locations of the
  Seyfert~2 and LINER galaxies from Rola \etal (1997).  The arrows are
  as in Fig.~1 except that the magnetic field for the top one is
  2\,$\mu$G\,cm$^{3/2}$}
\end{figure} 

\begin{figure}[h]
\vskip -0.3truecm
\centerline{
\psfig{figure=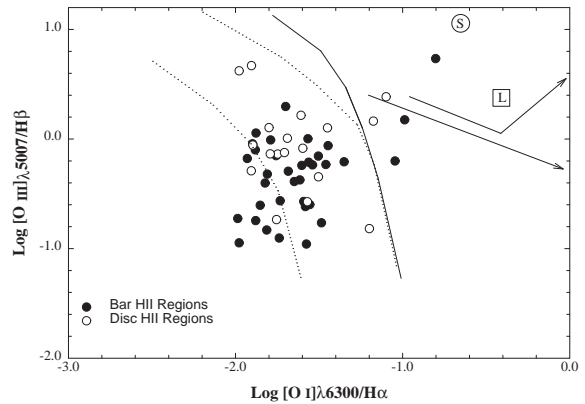,width=8.8cm,clip=}
}
\vskip -0.5truecm
\caption[]{Diagnostic diagram of bar and disc \hii\ regions. The
  dotted lines indicate the sequence of normal \hii\ regions. The full
  line is the separation between normal and abnormally excited regions
  (Osterbrock 1989).  The square and circular symbols correspond to
  the average location of Seyfert~2 and LINER galaxies from Rola \etal
  (1997).  The arrows are as in Fig.~1}
\end{figure}     

%--------------------------------------------------------------
\subsection{Electronic density}
The electronic density $N_e$ in \hii\ regions can be accessed with the
line ratio of the \sii\ doublet at 6717--6731\,{\AA}, i.e.  $\gamma
\!=\! {\sii\,\lambda 6717 \over \sii\,\lambda 6731}$
(Osterbrock 1989).  The density can be derived from nebular models
published by Blair \& Kirshner (1985).  Since the calibration depends
on the nebular temperature ($N_e \propto T_e^{0.5}$), we will assume
$T_e \!=\! 10^4$\,K.  This value is probably too high for the \hii\ 
regions in our sample with abundances higher than the solar value (see
Sect.~3.5).  However, for comparison purposes, this approximation is
appropriate.  The electronic density distributions for the bar and
disc \hii\ regions are illustrated in Fig.~4.  In both samples,
several regions show line ratios that are very close or larger than
the low-density limit ($\gamma \ga 1.4$).  The electronic densities
derived in this regime are very uncertain because the doublet ratio
becomes only weakly dependent on $N_e$ for values larger than about
1.3.

Kennicutt \etal (1989) found that nuclei \hii\ regions tend to possess
higher electronic densities on average than disc regions, with both
classes showing a large range of densities.  As seen in Fig.~4, the
case of \hii\ regions located in bars is different.  {\it No
  significant difference is observed between the distributions of
  electronic densities of both populations}.  On average, $\gamma
\approx 1.33 \pm 0.02$ (bar regions) and $\gamma \approx 1.31 \pm
0.03$ (disc regions).  Bar \hii\ regions do not show any compactness
with respect to disc regions.

%-----------------------------------------------------------
\begin{figure}[t]
\centerline{
\psfig{figure=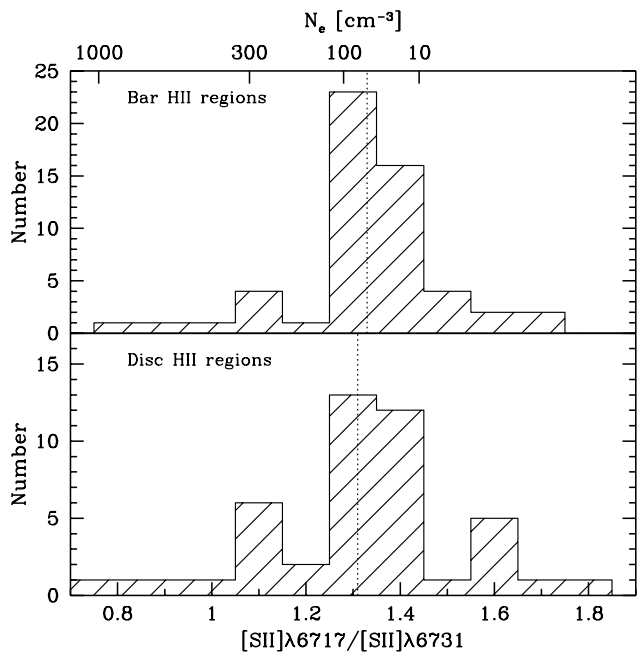,height=8.8cm,clip=}
}
\caption[]{Electronic density distributions for bar (top) and disc
  (bottom) \hii\ regions.  The density conversion derived from the
  \sii\ doublet ratio is from Blair \& Kirshner (1985). Vertical
  dotted lines indicate the mean values}
\end{figure}    

%-----------------------------------------------------------
\subsection{Extinction}
As discussed in the Sect.~2.2, the visual interstellar extinction of
individual \hii\ regions can be derived from the \hahb\ line ratio.
The values derived, however, are approximate since in reality, the
real extinction is probably not distributed uniformly but is patchy.
Figure~5 presents the distribution of the visual interstellar
extinction from both the bar and disc \hii\ regions.  There is a
considerable extinction in both populations of \hii\ regions.  The
mean values are $A_V \approx 1.3$ and $A_V \approx 1.0$ visual
magnitudes for the bar and disc regions, respectively.  This
difference is probably not significant since the extinction derived in
bar regions, based on the assumption that the nebular temperature is
$10^4$\,K, might be overestimated by about 0.1 to 0.2 magnitude.

%-----------------------------------------------------------
\begin{figure}[t]
\centerline{
\psfig{figure=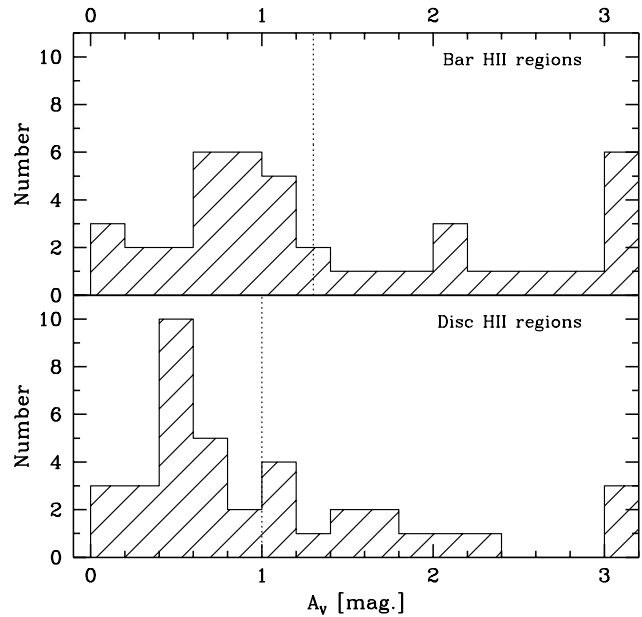,height=8.8cm,clip=}
}
\caption[]{Distributions of the interstellar extinction $A_V$ derived
  from the \hahb\ line ratio for bar (top) and disc (bottom) \hii\ 
  regions. Vertical dotted lines indicate the mean values}
\end{figure}    

%-----------------------------------------------------------
The extinction values for the bar regions show a large dispersion.
Most of the \hii\ regions contributing to the highest values are
located in the bars of NGC\,7479 and NGC\,5921.  As noticed in MF97,
these regions are located close to the strong dust lanes seen in these
bars.  Nevertheless, the overall behavior shows that the interstellar
extinction as derived from the Balmer decrement is similar for bar and
disc regions.  However, using IRAS observations, Phillips (1993) has
shown that for circumnuclear regions the extinction derived from the
\hahb\ ratio can be underestimated by as much as 2 magnitudes.  The
``uniform screen'' model assumed for the extinction is probably
over-simplistic.  Any line ratios (e.g. $R_{23}$) or other
quantitative properties (e.g. integrated fluxes) severely affected by
the interstellar extinction should be interpreted with caution for
regions located in the inner parts of galaxies.

%-----------------------------------------------------------
\subsection{\Ha\ equivalent widths}
The equivalent widths (EW) of the Balmer emission-lines provide a
measure between the number of ionizing and continuum photons emitted
in the \hii\ region.  As such, the EWs depend strongly on the stage of
evolution of the ionizing stars, the initial mass function (IMF), and
the metallicity (Dottori 1981; McCall \etal 1985; Copetti \etal 1986;
Bresolin \& Kennicutt 1997; Bresolin \etal 1999; Leitherer \etal
1999). As shown by Copetti \etal (1986) and more recently by Leitherer
\etal (1999), EW(\Ha) and EW(\Hb) can both be used as age indicators
for \hii\ regions. In disc galaxies, the distribution of EW(\Ha)
extends from about 100\,{\AA} to 1500\,{\AA} with a median value
around 400\,{\AA}.  No obvious correlation with the Hubble type is
found (Bresolin \& Kennicutt 1997).  Assuming an instantaneous burst
of star formation and a solar metallicity, these values correspond to
an age range between 1\,Myr to 7\,Myr (Leitherer \etal 1999).

The accuracy of the Balmer line EWs is mostly determined by the
uncertainty in the level of the nebular continuum which is severely
contaminated by the galactic continuum.  Because we could directly
measure the contribution from the galactic continuum on our ``off''
band images used in MF97, we have only measured the EWs for the \Ha\ 
line. The EW(H$\alpha$) is also less affected by the interstellar
extinction and the underlying absorption.  The fraction of the galaxy
light contributing to the nebular continuum was estimated from two
photometric apertures: one covering the integrated light of the \hii\ 
regions, and the other located on nearest area devoid of any \Ha\ 
emission (determined from the \Ha\ images).  A correction factor was
then applied to the EW values measured directly from the spectra.
These correction factors vary from 1.1 to 20 depending on the location
of the \hii\ region.

%-----------------------------------------------------------
\begin{figure}[t]
\centerline{
\psfig{figure=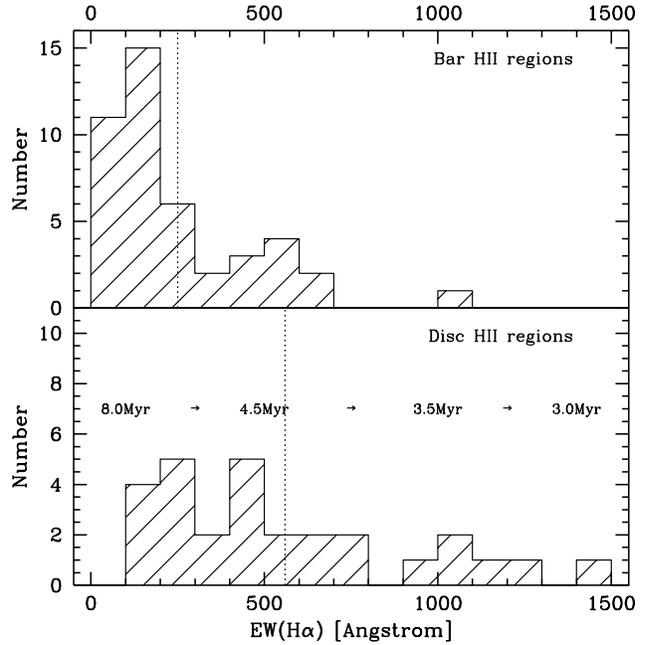,height=8.8cm,clip=}
}
\caption[]{Distribution of the \Ha\ equivalent widths EW(\Ha) for the
  bar (top) and disc (bottom) \hii\ regions.  The scale in the bottom
  panel indicates the approximate age of the \hii\ regions from models
  with a Salpeter IMF, instantaneous burst of star formation and solar
  metallicity (see Leitherer \etal 1999 for details).  Vertical dotted
  lines indicate the mean values}
\end{figure}    

%-----------------------------------------------------------
Figure~6 illustrates the EW(\Ha) distributions of the bar and disc
\hii\ regions.  The mean values for the distributions differ by about
a factor of two: EW(\Ha) $\approx$ 250\,{\AA} (bar regions) and
EW(\Ha) $\approx$ 560\,{\AA} (disc regions).  Following the models of
Leitherer \etal (1999) for an instantaneous burst of star formation
with a Miller-Scalo mass function and solar metallicity, the mean age
of bar regions is about 5.3\,Myr while disc regions are about 4.0\,Myr
old.  However, the difference in age is less ($<1 \times 10^6$\,yr)
when a Salpeter function is used to describe the IMF. Also, no 
age gradient seems to exist along the sequence of bar HII 
regions. This is an indication that HII regions should be ignified 
all along the bar and not only at bar ends with a subsequent 
 migration towards the center.

In their study of nuclear \hii\ regions, Kennicutt \etal (1989) found
that the EW(\Ha) of the \hii\ region nuclei, with a median value
around 25\,{\AA}, are approximately 20 times lower on average than
that of normal disc regions.  Such a difference cannot easily be
explained by assuming that the correction for galactic continuum was
underestimated.  The authors rather favor the idea that the stellar
continuum is high due to continuous star formation in the same region
or an unusual stellar mass spectrum in the ionizing clusters.  In the
present case, however, it is difficult to completely discard the
effect of the contamination of the galactic continuum to the nebular
continuum to explain the discrepancy observed between bar and disc
regions.  The location of the aperture used to measure the galaxy
continuum has a strong effect on the correction performed. The light
distribution in bars is not uniform and some bars have very patchy
dust features.  A systematic error of a factor 2 cannot be ruled out.
In any case, no firm conclusion based on the difference observed in
the EW({\Ha}) of both populations of \hii\ regions can be drawn from
the actual sample.

%-----------------------------------------------------------
\subsection{O/H abundance (within bars)}
The oxygen abundance in \hii\ regions can be derived either through
semi-empirical calibrations (e.g. Edmunds \& Pagel 1984; McGaugh 1991;
Pagel 1997) or directly when the temperature can be measured from the
nebular lines \oi\,$\lambda$4363 or \nii\,$\lambda$5755.  The latter,
however, are generally detectable only for \hii\ regions with low
oxygen abundance.  In our case, almost all the regions have solar or
above-solar oxygen abundances; semi-empirical techniques have to be
used.  Even if the uncertainties related to these methods are
generally quite large ($\pm$0.2\,dex), it is worthwhile to derive the
O/H values to address the important question of mixing of the ISM in
bars.  It is now well established that bars induce large-scale mixing
of the chemical composition in the disc of spirals (Martin \& Roy
1994; Zaritsky \etal 1994; Friedli \etal 1994).  The radial flows of
gas formed by bars flatten the strong (negative) abundance gradients
generally observed in unbarred late-type disc galaxies.  The
importance of the homogenization effect is related to the bar strength
as shown by Martin \& Roy (1994).  Very strong gas flows ($v \ga
100$\,km\,s$^{-1}$) are taking place along bars; efficient mixing
should be also observed and the O/H scatter between the bar \hii\ 
regions should be smaller than what is observed in normal galaxy discs
(See Sect.~4).

%-----------------------------------------------------------
\begin{figure}[t]
\centerline{
\psfig{figure=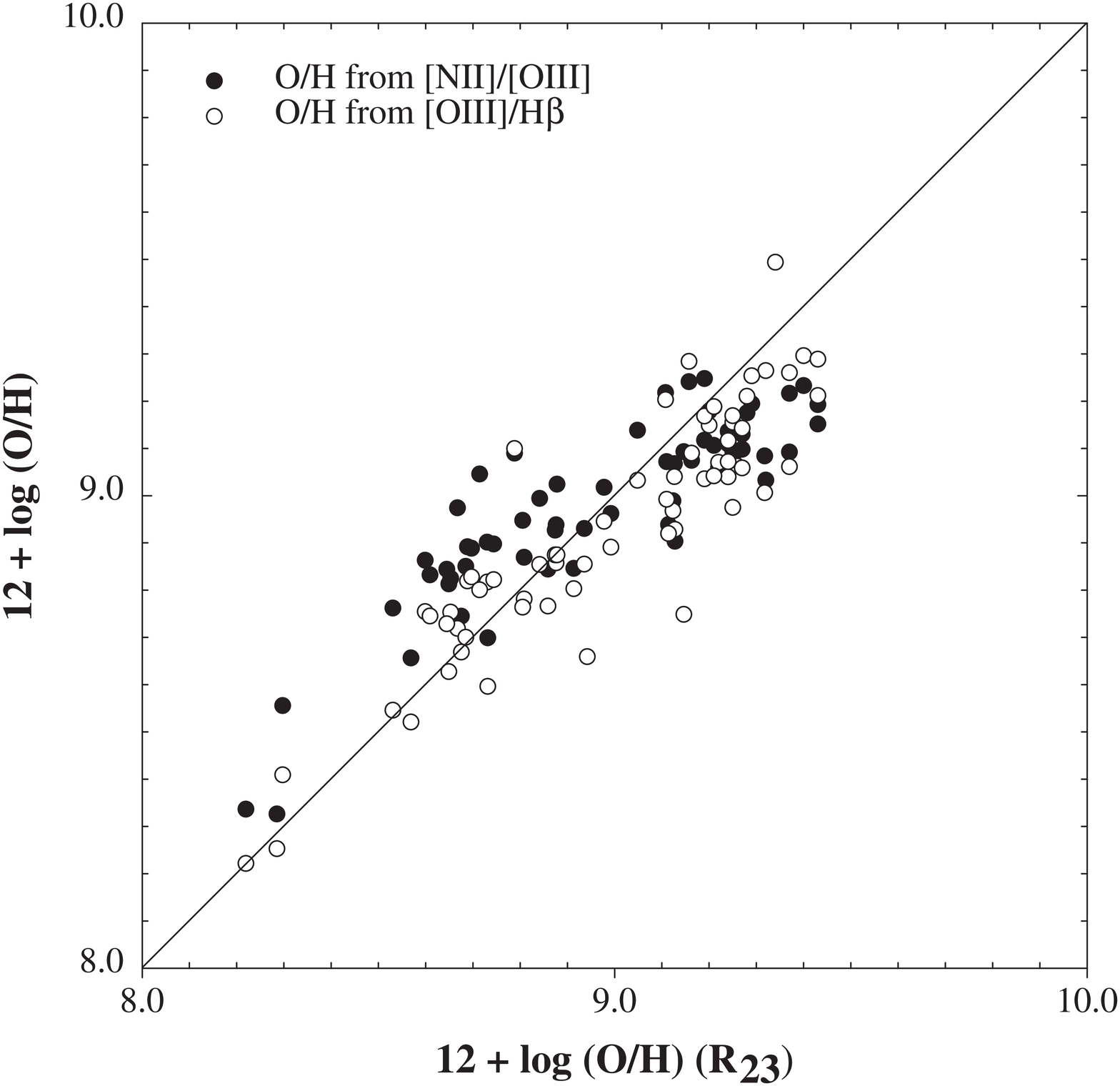,height=9.6cm,clip=}
}
\caption[]{Comparison between the O/H abundances values given by three
  different line ratios: \niioiii, \ohb, and $R_{23}$}
\end{figure}

%-----------------------------------------------------------
The oxygen abundances for our sample of bar \hii\ regions were
determined using three line ratios: \niioiii, \ohb, and $R_{23} \!=\!
{\oii\,\lambda 3727 + \oiii\,\lambda\lambda 4959,\,5007 \over \Hbm}$.
The conversion to relative oxygen abundances was done using the
calibration of Edmunds \& Pagel (1984).  Much has been written on the
accuracy of these line ratios as abundance indicators (e.g.  McGaugh
1991; Martin \& Roy 1995; Stasi\'nska 1998).  For our sample, Fig.~7
compares the different O/H values derived with all three indicators.
The O/H values derived from \niioiii\ are slightly higher
($\sim$0.1\,dex) than the values derived from \ohb\ and $R_{23}$ for
$12+\log(\rm O/H) < 8.9$.  For $12 + \log(\rm O/H) > 8.9$, that is,
the abundance regime for most of the bars in our sample, the
abundances from \ohb\ and \niioiii\ are slightly below the values
given by $R_{23}$.  These results were previously discussed by Martin
\& Roy (1994) and are due to discrepancies in the semi-empirical
calibrations.  For our purposes, we use the O/H values derived from
the \niioiii\ indicator; our conclusions are not affected by this
choice.

The distribution of the oxygen abundance along the bars of our sample
is illustrated in Fig.~8.  It is clear that {\it the O/H scatter
  observed in these bars is well smaller than $\pm$0.1\,dex or even
  less}.  Martin \& Belley (1996, 1997) have shown that the azimuthal
O/H dispersion observed in the discs of normal and barred galaxies is
generally equal or larger than $\pm$0.2\,dex.  These abundance
variations are probably not intrinsic but are in fact a combination of
real inhomogeneities and uncertainties associated with empirical
techniques (Roy \& Kunth 1995).  However, a comparative analysis
remains possible when the same method is used to derive the O/H values
(Martin \& Belley 1997).  Clearly, the chemical composition of \hii\ 
regions in a bar is well-homogenized.  This indicates that an
efficient mixing of the chemical composition is taking place in the
bar region (see next section).

%-----------------------------------------------------------
\begin{figure*}
\vskip -2truecm
\centerline{
\psfig{figure=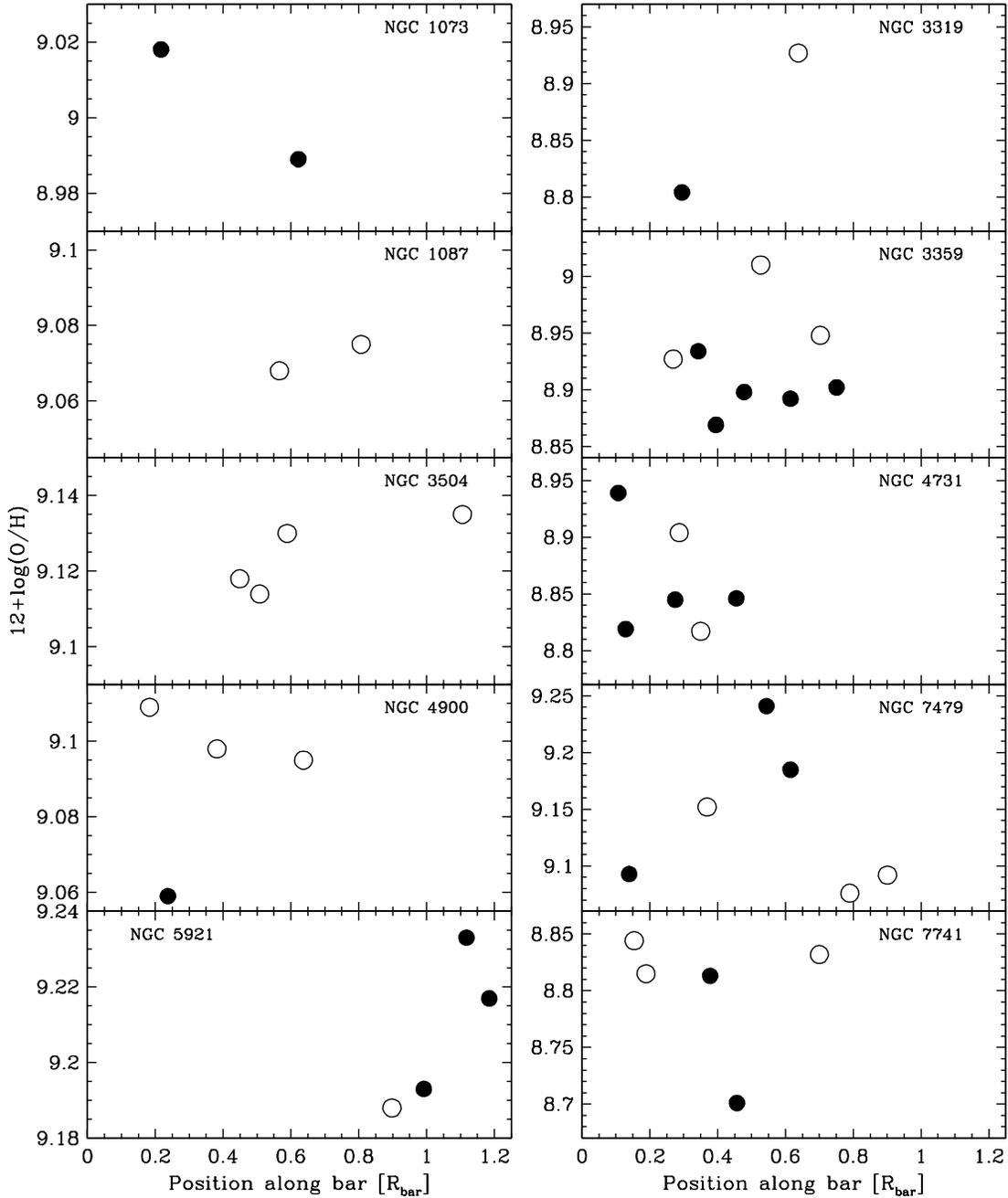,width=16cm,clip=}
}
\vskip -1.5truecm
\caption[]{O/H abundance distributions along the bars of the galaxies 
  in our sample.  The O/H values were derived using the \niioiii\ line
  ratio.  The horizontal axis is the position of the regions
  normalized to the bar radius given in MF97. The open and black
  symbols differentiate the regions with regard to what side of the
  bar they are located. The abundance interval is 0.06\,dex for
  galaxies displayed in the left column and 0.2\,dex for those in the
  right column}
\end{figure*}

%-----------------------------------------------------------
%
%  DISCUSSION
%  
\section{Discussion}
%-----------------------------------------------------------
{\it Degree of ionization.}  As discussed in Sect.~3.1, bar \hii\ 
region spectra do not exhibit any obvious signs of high-velocity 
shocks or hard
UV radiation.  However, there is marginal evidence from Figs.~2 and
3 that the ionization might be different for some bar regions. The
degree of ionization at a specific position in a nebula can be
accessed through the ionization parameter:
\begin{equation}
U = {Q(H^0) \over 4 \pi r^2 c N_e}
\end{equation}
where $Q(H^{0})$ is the number of ionizing photons per unit time by
the central source, $r$ is the position in the nebula, and $c$ the
speed of light (Osterbrock 1989). The most obvious evidence that bar
regions are different from disc regions is seen in the
\oi\,$\lambda$6300 line. As described in Evans \& Dopita (1985), the
\oi\ line is emitted in the transition zone of an \hii\ region which
contains a significant fraction of neutral hydrogen. The line is then
stronger when $U$ is lower (or the ionizing stellar temperature is
lower).  Since we have detected the \oi\ line in a much larger
fraction of bar regions than in disc regions, this suggests that the
ionization parameter could be lower in the former population.
Figure~9 shows the correlation between
\oi\,$\lambda$6300/\oiii\,$\lambda$5007 and
\oii\,$\lambda$3727/\oiii\,$\lambda$5007 which is particularly
dependent on $U$ (Evans \& Dopita 1985).  A correction for the
interstellar extinction has been applied to these line ratios.  The
bulk of bar \hii\ regions is located at $U$$\sim$0.0005.  For the disc
regions, the scatter is quite large but on average $U$$\sim$0.001,
larger than the value for bar regions.  Unfortunately, our sample is
not large enough to firmly confirm that $U$ is indeed different for
both populations of \hii\ regions.

%-----------------------------------------------------------
\begin{figure}[t]
\centerline{
\psfig{figure=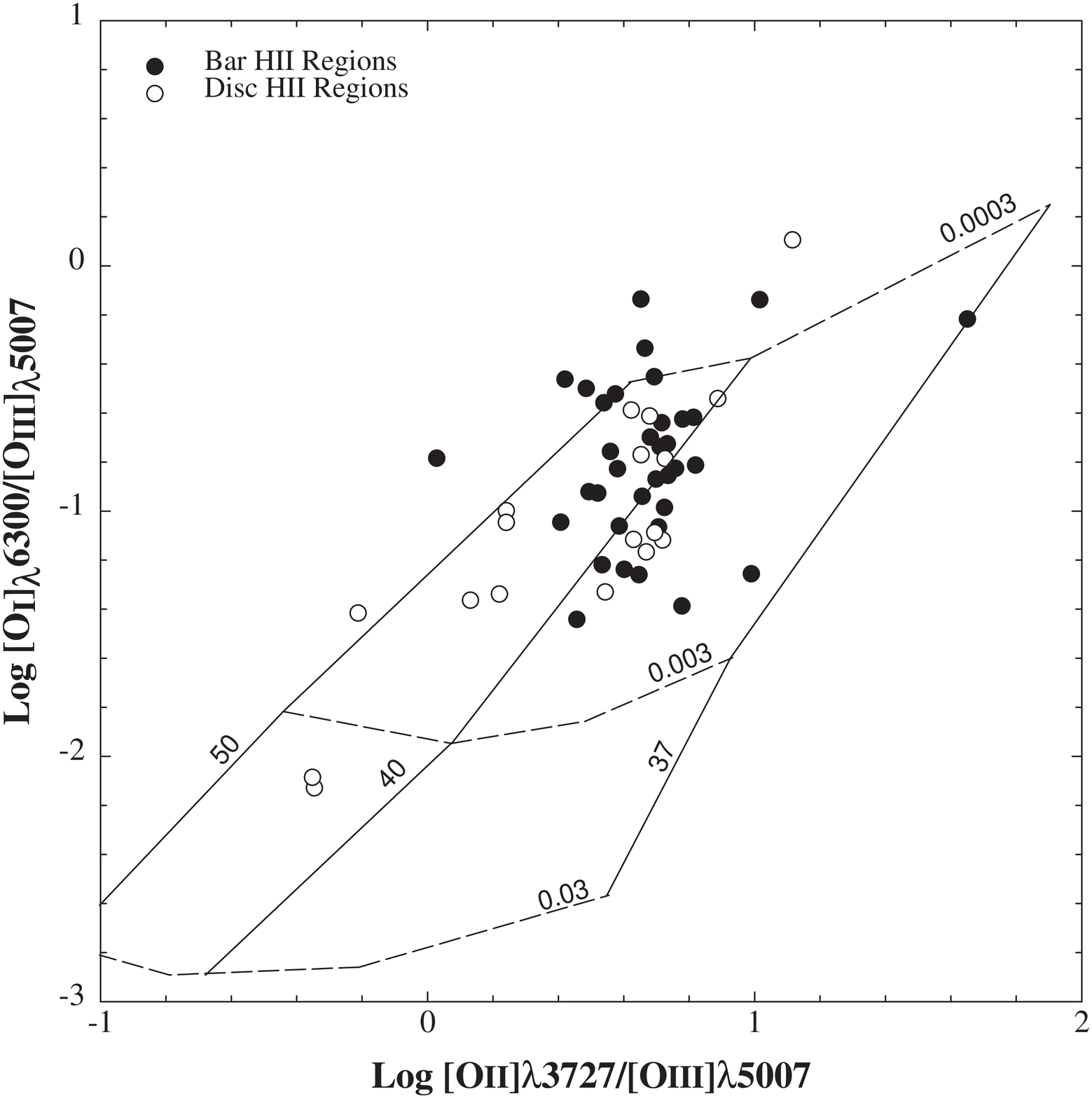,height=9.6cm,clip=}
}
\caption[]{Diagnostic diagram showing the relation between the 
  \oi\,$\lambda$6300/\oiii\,$\lambda$5007 and
  \oii\,$\lambda$3727/\oiii\,$\lambda$5007 nebular line ratios. Both
  ratios have been corrected for the extinction.  The curves represent
  models from Evans \& Dopita (1985).  The horizontal dashed lines
  show three values for $U$, 0.03, 0.003, and 0.0003.  The vertical
  lines indicate the stellar temperatures used in the models,
  50\,000\,K, 40\,000\,K and 37\,000\,K.}
\end{figure} 

%-----------------------------------------------------------
If a difference in the ionization parameter is really present, this
could be due to many factors: differences in the initial mass
function, age, richness of the OB associations, or spatial
distribution of the ionized material (Evans \& Dopita 1985). As
recently shown by Rozas et al. (1999), the luminosity function (LF) of
the bar regions is much less regular than the LF of the disc regions
in the strongly barred spiral NGC\,7479.  Their result, combined with
our study on the nebular excitation, suggest strongly that the
properties of the OB associations formed in bars differ from the
normal associations of the disc.  More work comparing LFs and nebular
properties of a larger sample of bar and disc regions would allow us
to investigate the origin of this difference.  In the final paper in
this series (Friedli \& Martin, in preparation), we will also examine
the properties of the clusters formed in diverse bar environments with
high-spatial numerical simulations.

\medskip
\noindent 
{\it Mixing and element production.}  The estimated timescale given by
numerical simulations for which the star formation activity phase
lasts in bars is $\tau_{\rm SF} \sim 5 \times 10^8$\,yr (Martin \&
Friedli 1997).  How does this compare with mixing timescale?

It is possible to roughly quantitatively evaluate the timescale of
mixing of the ISM due to radial flows.  Roy \& Kunth (1995) have
discussed the diverse mixing mechanisms of the oxygen abundance in the
ISM in galaxy discs.  Assuming a pure radial mixing due to gas flows
funnelled in the bar, the upper limit for the time for gas to diffuse
a length scale, $\Delta x_{\rm rad}$, in the radial direction is:
\begin{equation}
\tau_{\rm rad}={\Delta x_{\rm rad}^2 \over v l} \, , 
\end{equation} 
where $v$ is the radial flow velocity, and $l$ is the mean free path
for molecular clouds. In a typical bar, $\Delta x_{\rm rad} \!=\!
5$\,kpc (see Table 1) and $v\!=\!100$\,km\,s$^{-1}$.  The value of the
mean free path for the gas clouds is not a well-defined quantity in
bars.  In galaxy discs, $l\!=\!300-1000$\,pc (Roberts \& Hausman 1984;
Roy \& Kunth 1995). If we assume $l\!=\!500$\,pc for bars, we find
$\tau_{\rm rad} \approx 5 \times 10^8$\,yr.  However, since radial
flows in bars are not stationary, the real mixing timescale could be
even of the order of $\tau_{\rm rad} \!=\! \Delta x_{\rm rad}/v$, i.e.
$\sim 5 \times 10^7$\,yr. Putting all this together yields the
following reasonable interval for the mixing timescale: $5 \times 10^7
\la \tau_{\rm rad} \la 5 \times 10^8$\,yr.  Thus, $\tau_{\rm rad}$ is
shorter than $\tau_{\rm SF}$ meaning that the abundance content in
\hii\ regions formed during this phase must be homogenized.

It is also instructive to make rough (i.e. close-box) estimates of
the global abundance increase during $\tau_{\rm SF}$ as well as of the
abundance fluctuations in \hii\ regions which should result from their
age spread. If $Z^i$ is the initial mean gaseous abundance in the bar
region, $\epsilon$ the global star formation efficiency, and $Y_Z$ the
net yield for the species considered, then the final mean abundance is
given by: 
\begin{equation} 
Z^f = Z^i + \epsilon (1-\epsilon)^{-1} Y_Z \, .  
\end{equation}
Interestingly enough, $Z^f$ does not depend on the initial gas mass
fraction. For instance, for the oxygen with an yield $Y_O \approx
0.006$, an initial solar abundance $Z^i \approx 0.01$, and a typical
SF efficiency $\epsilon \approx 0.25$, then $Z^f \approx 0.012$.  The
global increase of oxygen abundance is thus only about 0.1\,dex.
Equation~3 can in fact also be applied to each individual \hii\ 
regions with exactly the same numbers; the fluctuations in the oxygen
abundance are thus expected to be of the order 0.1\,dex, which is
indeed what is observed (Sect.~3.5).  However, we do not observe any
clear trend between the age and metallicity for bar \hii\ regions.  In
dwarf galaxies, the metal enrichment of \hii\ is not observed and
metals are probably locked in the hot phase.  (see e.g. Tenorio-Tagle
1996; Kobulnicky 1998).  The situation could be similar for star
forming regions in bars.

%---------------------------------------------------------------------
%
%  SUMMARY
%
\section{Summary}
%-----------------------------------------------------------
The main results concerning the properties of \hii\ regions located
within bars can be expressed as follows:

\smallskip
\noindent
{\it 1) From standard diagnostic diagrams, the excitation of most
  \hii\ regions appears normal and similar to the one of disc
  regions.} There are some exceptions for nuclear regions where an
ionization mechanism other than photoionization seems to be present.
However, there is marginal evidence that the ionization parameter in
bar regions is lower than in disc regions, suggesting that the
properties of the OB associations might be different in bar
environments.

\smallskip
\noindent
{\it 2) The electronic density distribution as derived from the \sii\ 
  line ratio is similar to that observed for normal disc regions.}
The mean density is $N_e \approx 80$\,cm$^{-3}$.  Star formation
regions in bars have the same ``compactness'' as disc regions.

\smallskip
\noindent
{\it 3) The \hahb\ extinction indicator reveals that, on average, bar
  regions have a visual extinction $A_V \sim 1.3$\,mag, 0.3\,mag more
  than disc regions.} This difference is mainly due to the fact that
some regions are located near the bar dust lanes of the earlier types
of galaxies in our sample.

\smallskip
\noindent
{\it 4) The average \Ha\ equivalent width for the bar regions is about
  250\,\AA, half that of disc regions.} While this could indicate an
older population for bar \hii\ regions, the corresponding age
difference is probably too small to be significant owing to the large
uncertainties introduced by the galactic continuum correction.

\smallskip
\noindent
{\it 5) The O/H abundance distribution of these \hii\ regions is
  remarkably homogeneous.}  This is the result of the gaseous radial
flows in bars inducing mixing of the ISM, as seen on a larger scale in
the discs of barred spirals.

%----------------------------------------------------------------------------
\begin{acknowledgements}
  Discussions with P.~Ferruit, L. ~Binette, R.~Kennicutt, and 
  J.-R.~Roy were most 
  appreciated throughout this work.  We also thank the referee, Fran\c{c}oise 
  Combes, for her helpful comments. The efficient support 
  offered by the technical staff of
  the Multiple Mirror Telescope during the observations was much
  appreciated. This work was supported by NSERC (Canada), FCAR
  (Qu\'ebec) and in part by the NSF through grant AST-94-21145.  D.F.
  acknowledges the kind hospitality of CFHT.
\end{acknowledgements}

%----------------------------------------------------------------------------

%----------------------------------------------------------------------------

\begin{thebibliography}{}
  
\bibitem{} Baldwin J., Philips M., Terlevich R., 1981, PASP 93, 5

\bibitem{} Blair W.P., Kirshner R.P., 1985, ApJ 289, 582

\bibitem{} Bresolin F., Kennicutt R.C., 1997, AJ 113, 975
  
\bibitem{} Bresolin F., Kennicutt R.C., Garnett D.R., 1999, ApJ 510,
  104
  
\bibitem{} Copetti M.V.F., Pastoriza M.G., Dottori H.A., 1986, A\&A
  156, 111

\bibitem{} Dopita M.A., Sutherland R.S., 1995, ApJ 455, 468

\bibitem{} Dottori H. A., 1981, Astrophys. Space Sci. 80, 267

\bibitem{} Edmunds M.G., Pagel B.E.J., 1984, MNRAS 211, 507 

\bibitem{} Evans, I. N, Dopita, M. A., 1985, ApJS 58, 125

\bibitem{} Friedli D., Benz W., 1995, A\&A 301, 649
  
\bibitem{} Friedli D., Benz W., Kennicutt R.C., 1994, ApJ 430, L105
  
\bibitem{} Garc\'{\i}a--Barreto J.A., Franco J., Carrillo R., Venegas
  S., Esca\-lante-Ram\'{\i}rez B., 1996, RevMexAA 32, 89
  
\bibitem{} Kennicutt R.C., Keel W.C., Blaha C.A., 1989, AJ 97, 1022
  
\bibitem{} Kobulnicky H.A., 1998, in Abundance Profiles: Diagnostic
  Tools for Galaxy History, ASP Vol.~147, eds. D.\,Friedli et al. (San
  Francisco: ASP), 108
  
\bibitem{} Leitherer C., Schaerer D., Goldader J.D., et al., 1999,
  ApJS, in press

\bibitem{} Martin P., Belley J., 1996, ApJ 468, 598 

\bibitem{} Martin P., Belley J., 1997, A\&A 321, 363

\bibitem{} Martin P., Friedli D., 1997, A\&A 326, 449 (MF97)

\bibitem{} Martin P., Roy J.-R., 1994, ApJ 424, 599 

\bibitem{} Martin P., Roy J.-R., 1995, ApJ 445, 161 

\bibitem{} Martinet L., Friedli D., 1997, A\&A 323, 363

\bibitem{} McCall M.L., Rybski P.M., Shield G.A., 1985, ApJS 57, 1

\bibitem{} McGaugh S.S., 1991, ApJ 380, 140
  
\bibitem{} Osterbrock D.E., 1989, Astrophysics of Gaseous Nebulae and
  Active Galactic Nuclei (Mill Valley: Univ. Science Books)
  
\bibitem{} Pagel B.E.J., 1997, Nucleosynthesis and Chemical Evolution
  of Galaxies, (Cambridge: Cambridge Univ. Press)
  
\bibitem{} Phillips A.C., 1993, Ph.D. Thesis, Univ. of Washington, USA
  
\bibitem{} Phillips A.C., 1996, in Barred Galaxies, ASP Vol. 91, eds.
  R.\,Buta et al. (San Francisco: ASP), 44

\bibitem{} Roberts W.W., Hausman M.A., 1984, ApJ 277, 744 
  
\bibitem{} Rola C.S., Terlevich E., Terlevich R.J., 1997, MNRAS 289,
  419
  
\bibitem{} Rozas M., Zurita A., Heller C.H., Beckman J.E., 1999,
  A\&AS 135, 145

\bibitem{} Roy J.-R., Kunth D., 1995, A\&A 294, 432

\bibitem{} Stasi\'{n}ska G., 1990, A\&AS 83, 501 
  
\bibitem{} Stasi\'{n}ska G., 1998, in Abundance Profiles: Diagnostic
  Tools for Galaxy History, ASP Vol.~147, eds. D.\,Friedli et al. (San
  Francisco: ASP), 142

\bibitem{} Savage B.D., Mathis J.S., 1979, ARA\&A 17, 73  

\bibitem{} Tenorio-Tagle G., 1996, AJ 111, 1641 

\bibitem{} Veilleux S., Osterbrock D.E., 1987, ApJS 63, 295
  
\bibitem{} Zaritsky D., Kennicutt R.C., Huchra J.P., 1994, ApJ 420, 87

\end{thebibliography}
\end{document}